\documentclass{sig-alternate}
\usepackage{algorithm}
\usepackage{algorithmic}
\usepackage{amsmath}
\usepackage{amsfonts}

\usepackage{balance}
\makeatletter
\def\@copyrightspace{\relax}
\makeatother

\title{Distributed TensorFlow with MPI}
\author{%
Abhinav Vishnu{\small $~^{\#1}$}
Charles Siegel{\small $~^{\#2}$}, and
Jeff Daily{\small $~^{\#3}$}
\vspace{1.6mm}\\
\fontsize{10}{10}\selectfont\itshape
$^{\#1,2,3}$\,Pacific Northwest National Laboratory,
Richland, WA 99352\\
}
\begin{document}
\maketitle

\begin{abstract}
Machine Learning and Data Mining (MLDM) algorithms are becoming
increasingly important in analyzing large volume of data generated by
simulations, experiments and mobile devices. With increasing data
volume, distributed memory systems (such as tightly connected
supercomputers or cloud computing systems) are becoming important in
designing 
in-memory and massively parallel MLDM algorithms. Yet, the majority of open
source MLDM software is limited to sequential execution with a few
supporting
multi-core/many-core execution.

In this paper, we extend recently proposed Google TensorFlow for
execution on large scale clusters using Message Passing Interface (MPI).
Our approach requires minimal changes to the TensorFlow runtime --
making the proposed implementation generic and readily
usable to increasingly large users of TensorFlow. We evaluate our
implementation using an InfiniBand cluster and several well known
datasets. Our evaluation indicates the efficiency of our proposed
implementation. 
\end{abstract}

\section{Introduction}
Today, simulations, experiments and mobile devices are generating
increasingly large volume of data~\cite{data:ascac11,data:ascac13}.
Machine Learning and Data Mining (MLDM) algorithms, which can build
models, classifiers, and anomaly detectors are being designed and
applied in several domains including high energy physics, computational
biology, and cyber security~\cite{tarca:bio07,vossen:hep08,ml:climate}.

MLDM algorithms are generally classified as {\em supervised} (the input
dataset is labeled with ground truth) and {\em unsupervised} (learning
from un-labeled dataset). Base unsupervised/supervised algorithms can be
combined together using {\em ensemble} methods to remove noise, and
possibly learn better models/classifiers. Several software packages
which support supervised, unsupervised and ensemble algorithms have been
released publicly. A few well known packages are Weka~\cite{weka},
Scikit~\cite{scikit}, libsvm~\cite{libsvm}, and Matlab. However, these
packages only support sequential execution. As a result, they are
generally used with modest size datasets.

At the same time, Deep Learning algorithms -- a class of MLDM algorithms
-- are becoming increasingly popular. Deep Learning algorithms emulate
brain activity by using several layers of neurons (interconnected with
synapses) and learn the weights for the synapses by using gradient
descent methods. There are several classes of Deep Learning algorithms
-- Deep Neural Networks (DNN - typically used on tabular datasets),
Convolutional Neural Networks (CNNs - typically used on images) and
Recurrent Neural Networks (RNNs - typically used on time-dependent
datasets).   Several researchers/practitioners have applied Deep
Learning algorithms to their problems, and reported better results in
comparison to their well published models. Naturally, open source
efforts such as Theano, CuDNN, and Caffe~\cite{caffe} have gained traction and wide
acceptance among researchers and practitioners alike.

Recently, Google released TensorFlow, which is a toolkit for developing
MLDM algorithms. It uses a dataflow model by specifying operations on tensors
(user-defined multi-dimensional arrays). It also supports automatic
differentiation, which simplifies the design and implementation of
gradient descent methods. TensorFlow readily supports DNNs, CNNs and
RNNs on multi-core/many-core systems (GPUs) and supports algorithmic
advancements such as AdaGrad, and Neuron Dropout for regularization. 
However, TensorFlow's restriction to single compute node is highly
restrictive, especially with increasing size of the datasets.

In this paper, we propose a design to alleviate these limitations of
TensorFlow. Specifically, we extend TensorFlow for scalable execution on
very large scale systems. We consider several programming models,
especially MapReduce based programming models (Hadoop, and Spark) and
conclude that neither of them are geared towards realizing the peak
potential of the system, while TensorFlow is geared towards exploiting the
architecture effectively using a C++ backend and state of the art linear
algebra packages. We use Message Passing Interface (MPI)~\cite{mpi1}
as the communication interface for parallelizing TensorFlow on
distributed memory subsystems. We specify the changes which were
required to realize the implementation on distributed memory systems.
Specifically, we conclude that these changes are minimal and require
no changes to the TensorFlow runtime! Our evaluation of the proposed
extensions with several well known datasets such as MNIST, CIFAR-10,
Adult and Higgs reveals the performance efficiency of the proposed
implementation.

\section{Background}
\label{sec:background}
In this section, we provide a brief background of Google TensorFlow
(simply referred as TensorFlow for rest of the paper) and Message
Passing Interface (MPI)~\cite{mpi1, mpi2}.
\subsection{TensorFlow}
Google's TensorFlow, released in November 2015, is a platform for
building and developing models in machine learning, particularly neural
networks.  It is capable of handling multiple threads and devices on a
single machine, including a heterogeneous environment consisting of a
multi-core CPU and potentially multiple GPUs.

The basic unit in TensorFlow is the computational graph.  This graph
contains nodes, which are operations, and edges which represent tensors
(arbitrary dimensional arrays).  Each node can take multiple inputs and
give multiple outputs, with tensors created and passed from one node to
another and, generically, deleted after use to avoid memory clutter.
In addition to carrying tensors, edges can also be used to control the
flow of a computation.  Control dependencies can be used to enforce
relationships such that some computations must be done before others, no
matter what parallelization has occurred.

TensorFlow uses two special types of tensors.  The first, placeholders,
are the only place that input can go into a graph.  Other than through
placeholders, there is no way that data can enter the graph.  So,
placeholders are used for both the initial input of the training data
and labels, as well as the validation and testing data used to determine
if an algorithm works.
The other special type of tensor that TensorFlow uses is the variable.
Variables are tensors that are stored, rather than deleted after use, by
the computational graph.  For our implementation, the weights of the model
are stored as variables, so that they can be updated throughout the
training as a running computation.

All graph computations take place within a session.  At the beginning of
a session the computational graph is empty and there are no variables.
The session interprets the commands of the user to initialize variables
and to build the computational graph.  Then, the session runs the
computational graph, and must be called whenever the user wants to
extract information, such as the value of a variable or the successful
prediction rate on the test set.

TensorFlow determines what order to compute the graph in by creating a
queue of nodes with no dependencies.  It keeps track of the number of
unresolved dependencies for each node, and whenever it drops to zero,
that node is put into the queue.  The program then executes the nodes in
the queue in some order, continuing to decrease the unresolved
dependencies until it has computed the whole graph.

Parallelization in TensorFlow is done in a task-based manner.  That is,
each node is assigned to a device for computation, rather than running
the whole graph, in parallel, on multiple devices.  The way that it
assigned is via a greedy algorithm.  First, TensorFlow runs a simulation
of the graph to determine approximately how long each node will take to
compute and to determine the computation order as above.  Then, the
greedy algorithm assigns nodes to devices based on whether or not there
is a kernel for that operation on that device (not all operations have
GPU implementations, for instance) and based on which device is expected
to be free when the computation is ready to be done.

Finally, TensorFlow inserts send and receive nodes between devices to
transfer the tensors.  It does this in a way to minimize communication
(given the assignment of the graph) and modifies the graph assignments
slightly if it changes the total execution time to change where
communication happens.

\subsection{Message Passing Interface}
Message Passing Interface (MPI)~\cite{mpi1,mpi2} provides a rich set of
abstractions for inter-process communication. MPI supports pair-wise
communication (such as using send, receive) and group communication
(such as using reduction, barrier). MPI has become the {\em de facto}
communication interface for legacy scientific applications. 

The primary reason for MPI's success is its wide availability. MPI is
available on large scale supercomputers, cloud computing systems and it
can also be used for inter-process communication on a single compute
node -- if other shared memory programming models are not available. 
Unlike other runtimes such as Spark and GRPC, MPI is able to take
advantage of high performance interconnects such as InfiniBand, Cray
interconnects and IBM Blue Gene interconnects readily. Due to the
performance reasons, we considered MPI to be the primary communication
interface instead of other communication subsystems. 

Specifically, we have used several MPI routines for our large scale
implementation. We have used All-to-all reduction (an MPI primitive
which allows operations such as sum on user's data, and disseminates the
final result among all the processes in a group) for averaging weights
and biases and point-to-point operations for data distribution.

We also observed that MPI has been criticized for its lack of support
for fault tolerance. However, with recent advancements -- such as
User-level Fault Mitigation (ULFM) -- and open source implementations,
it is possible to design fault tolerant MLDM algorithms using MPI,
without losing performance and "continued execution" in the presence of
hardware faults. We expect that with ULFM (or its variants) becoming
available with mainstream implementations, MPI would find its wide
acceptance in the MLDM community.

\section{Solution Space}
\label{sec:design}
In this section, we present a solution space for distributed TensorFlow.
We specifically consider several programming models, and other design
choices, such as making changes to the TensorFlow runtime.

\subsection{Programming Models Solution Space}
There are several programming models which may be used a distributed
memory implementation of TensorFlow. Specifically we considered several
Mapreduce implementations including Hadoop and Spark. Hadoop was
removed from consideration due to its frequent communication to the I/O
subsystem. Spark -- which considerably improves upon Hadoop by
in-memory execution -- was considered for distributed memory
implementation. However, the current implementation of Spark runtime
suffers from two primary issues: inability to take advantage of high
performance communication networks using native interfaces (such as
Verbs on InfiniBand, and PAMI on Blue Gene/Q networks); frequent I/O due
to saving the key-value pairs for fault tolerance reasons. 

We addressed the limitations of Spark by using MPI as the communication
interface. Since MPI is primarily geared towards supercomputers, most
MPI implementations use the communication interface natively as much as
possible. When a native communication interface is not available, MPI
implementations use sockets interface -- making them equivalent to
runtime implementations of Spark. 

The other issue is saving the intermediate state of the application
for fault tolerance purposes. We use ULFM for this purpose -- which
allows the MPI application to continue executing in the presence of
faults. By using data parallelism (the model is replicated on each
node to minimize intra-epoch communication), the critical data
structures are automatically replicated for fault tolerance. Using these
approaches, we are able to address the limitations of Spark.

\subsection{TensorFlow Runtime Solution Space} 
There are several design choices for parallelizing TensorFlow
computation graph. For implementation in distributed memory, one design
choice is to make changes to the TensorFlow runtime, such that the
details of the implementation are completely abstracted from the user.
However, this choice suffers from several drawbacks. Primarily, this
choice makes the implementation less compatible with the frequent
updates to the TensorFlow runtime -- as expected in the upcoming
releases. In addition, the overall engineering difficulty associated
with this approach is non-trivial.

\subsection{Proposed Design and Implementation}
We instead use an alternative approach for distributed memory
implementation. We primarily use TensorFlow backend as a blackbox and
leverage its primities to support distributed memory execution. 

\subsubsection{Work Distribution}
Firstly, we split the samples across all TensorFlow devices. While it is
possible to consider the split unequally (such as a GPU TensorFlow
device is much more computationally powerful in comparison to a single
compute box). In the current implementation, each device is considered
of equal compute capacity. We intend to address this limitation in the
upcoming releases of our code. 

In the current implementation, the default process (in MPI terminology,
the process with rank zero) reads the samples from the disk and splits
them across processes. While this implementation is not optimized for
parallel reading, we consider this to be a minor issue, since the
majority of time is spent in training the network. We will consider
other methods for parallel reading in upcoming releases of the proposed
implementation.

\subsubsection{Data Parallelism}
We considered several methods for parallelism. Firstly, we considered
the methods where the matrices belonging to each layer (neurons
connected with synapses) were distributed among multiple compute nodes,
possibly with block/row decomposition. However, this approach requires
significant communication for each sample -- hence other approaches were
considered.

Another possible approach for consideration is equivalent to the
DistBelief proposed by Dean {\em et al.}. Under that approach, each worker
(equivalent to a device in TensorFlow) may update the weights/biases on
a parameter server asynchronously. However, DistBelief suffers from
bottleneck at parameter server, especially at scale. In addition, if
each worker updates their parameters at the end of a batch/epoch, they
are likely to cause severe network bottleneck. Hence, this approach was
disregarded.  

Another training model supported by TensorFlow is by splitting a
TensorFlow graph among devices. The samples are then pipelined across
different devices for training. However, this approach does not scale
well, this is limited by the depth of the training network. In many
cases, the number of layers is three -- which makes this approach
invalid. 

To alleviate these limitations, we considered an approach where the
model is replicated on each device. Each device learns the model
independently using standard backpropagation algorithm. This approach
scales well in computation and communication, even though the model is
replicated on each device. To support this argument, let us consider a
simple performance model of computation and communication at each epoch
during the training process. 

Let $m$ be the number of samples, and $p$ be
the number of processes. For simplicity, let $n$ be the number of
neurons in each layer and $l$ be the number of layers. Hence, at each
epoch, the total number of FLOPs (floating point operations) is
$\frac{m}{p}\cdot n^2 \cdot l$, while the total communication volume is
$n^2 \cdot l$. Naturally, with strong scaling -- the work per device
reduces -- however for reasonable work distribution, the overall time in
communication can be managed. By using MPI and high performance
communications, the overall fraction of time spent during computation is
increased. Hence, we implement this form of parallelism for our
implementation. This approach is also referred to as {\em model
parallelism} for rest of the paper.

\subsubsection{Synchronous/Asynchronous Updates}
An important design choice is the synchronization of weights and biases
with data parallelism. Several researchers have considered asynchronous
methods for updating these data structures. While there are certain
advantages of asynchronous updates -- it becomes difficult to reason
about the correctness of the algorithm and its equivalence to the
standard gradient descent algorithms. 

Hence, we consider synchronous methods for updating the weights and
biases. There are several reasons that this approach scales with MPI and
with the presence of high performance interconnects. Since MPI is heavily
optimized for Supercomputers by using native communication interfaces,
the overall time spent in communication is much smaller in comparison to
using sockets interface, which involves multiple copies. Additionally,
the averaging operation for synchronizing the data structures is
heavily optimized in MPI. There are several well known algorithms, which
implement the All-to-all reduction operation in $\log(p)$ time. Other
interconnects such as Blue Gene and InfiniBand support these operations
in hardware -- further reducing the overall time complexity of the
proposed implementation.

\section{Experimental Evaluation}
\label{sec:exp}
In this section, we present an experimental evaluation of the proposed
approaches using an InfiniBand cluster. Each machine in the system
consists of a multi-core Intel Haswell CPU, and 64 GB RAM. The machines are
connected using InfiniBand. We use OpenMPI 1.8.3 for performance
evaluation.

\subsection{Data Sets and Network Architectures}
\label{subsec:dc}
We use several data sets for performance evaluation.
Specifically, we have used MNIST, CIFAR10, Adult,
Acoustic, and Higgs Boson data sets for comparing the performance. Since
MNIST and CIFAR10 are structured data sets, we have used DNN and CNN
for evaluating them. Table~\ref{sample-table} shows the network
architecture used for performance evaluation.

\begin{table}[t]
		\caption{Deep Learning Algorithms and Network Architectures
		used for Data Sets in this paper}
		\label{sample-table}
		\vskip 0.15in
		\begin{center}
				\begin{small}
						\begin{sc}
								\begin{tabular}{lcccr}
			\hline
			Data set & Algo & Network Architecture\\
			\hline
			Adult    & DNN& 123-200-100-2 \\
			Acoustic     & DNN& 50-200-100-3         \\
			MNIST    & DNN& 784-200-100-10         \\
			MNIST     & CNN& 32,64 (Conv), 1024 (Full)         \\
			CIFAR10     & DNN& 3072-200-100-10         \\
			CIFAR10    & CNN& 32,64 (Conv), 1024 (Full)         \\
			HIGGS    & DNN& 28-1024-2         \\
			\hline
	\end{tabular}
						\end{sc}
				\end{small}
		\end{center}
		\vskip -0.1in
\end{table}

For CNN, we use several convolution layers followed by fully
connected layers (without any complex branching as in
GoogLeNet\cite{43022}).  The convolution layers consist of $5\times5$ windows,
step size 1, and are ReLU neurons, and these are always followed by
max-pooling layers pooling $2\times 2$ blocks.  This part of the network
is followed by fully connected layers of sigmoid neurons, followed by a
softmax output layer.

\subsection{MNIST}
\label{subsec:mnist}
The MNIST database of handwritten numbers is a widely used data set in
Machine Learning. We consider MNIST-DNN and MNIST-CNN evaluation.

\subsubsection{MNIST-DNN}

\begin{figure}[ht]
\vskip 0.2in
\begin{center}
\centerline{\includegraphics[width=\columnwidth]{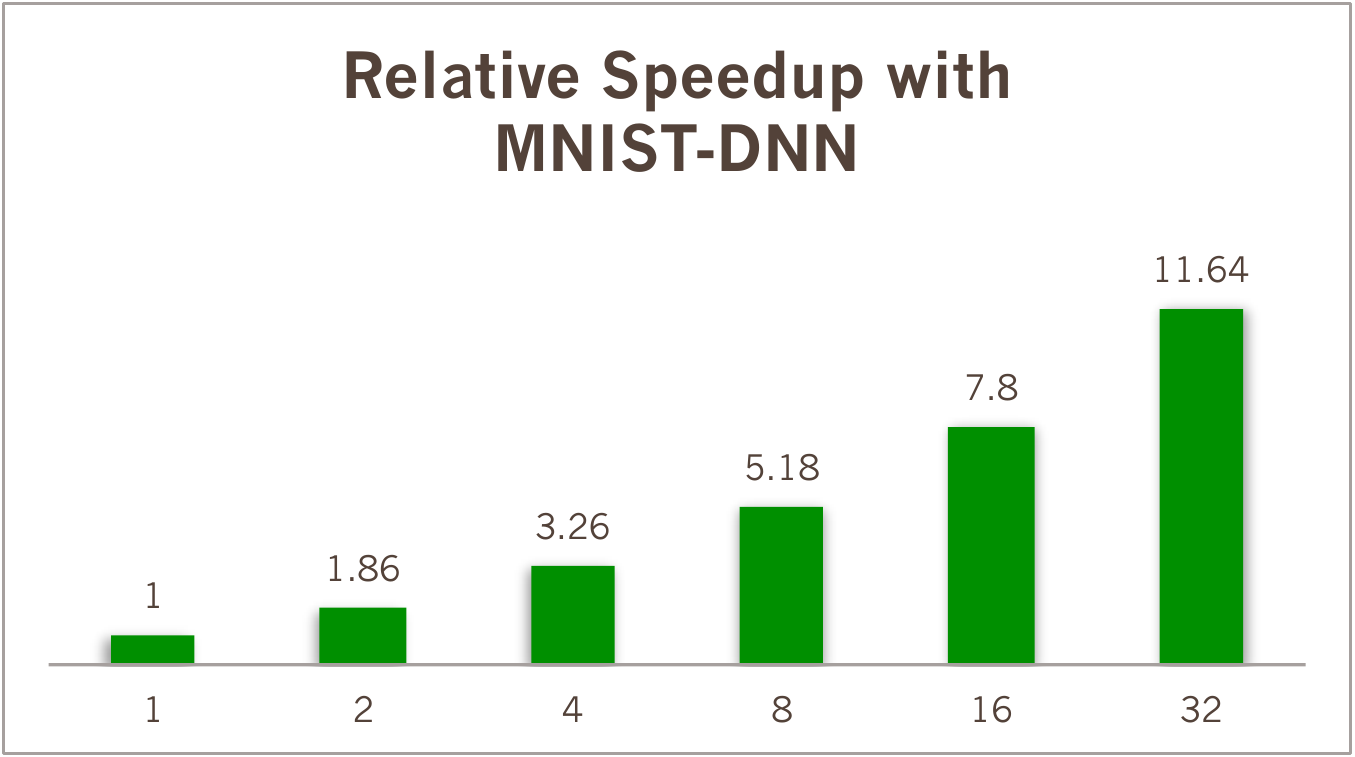}}
\caption{Relative Speedup to 1-core on MNIST-DNN using up to
32 cores}
\label{mnistdnn}
\end{center}
\vskip -0.2in
\end{figure}

Figure~\ref{mnistdnn} shows the relative speedup to
1-core with increasing number of cores. 

We observe a few trends from the chart: 
The approach scales well with increasing number of cores,
however, the overall improvement decreases due to strong scaling. 
We attribute this to the
decreasing amount of work available per core.
On smaller core counts -- such as available on desktops, the proposed 
approach would produce major performance improvement, as
shown in the chart. We also expect that with larger network
architectures, the relative improvement of this approach will
remain intact, since the overall work per core will increase. 
Overall we can achieve {\bf 11.6x} speedup.

\subsubsection{MNIST-CNN}
\begin{figure}[ht]
\vskip 0.2in
\begin{center}
\centerline{\includegraphics[width=\columnwidth]{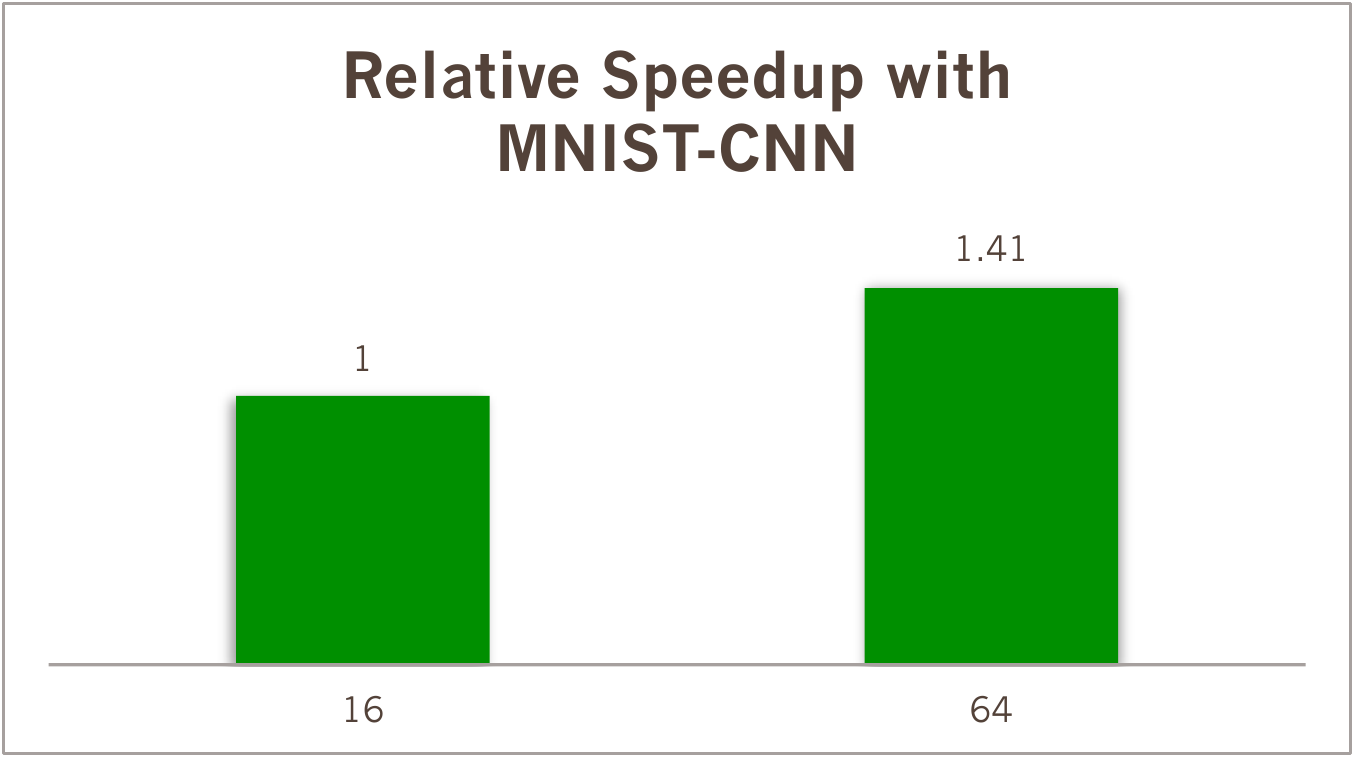}}
\caption{Relative Speedup to 16-core on MNIST-CNN using up to
64 cores}
\label{mnistcnn}
\end{center}
\vskip -0.2in
\end{figure}

Figure~\ref{mnistcnn} shows the relative speedup to 16-core
experiment. We observe that the improvement is up to 1.92x for 64 cores.
A factor which contributed
to the diminished improvement is that we trained the network for a fixed
time due to limited access to compute resources. We have observed that
with increasing the number of epochs, the benefits of the proposed approach increases.

\subsection{Adult} \label{subsec:adult}
\begin{figure}[ht]
\vskip 0.2in
\begin{center}
\centerline{\includegraphics[width=\columnwidth]{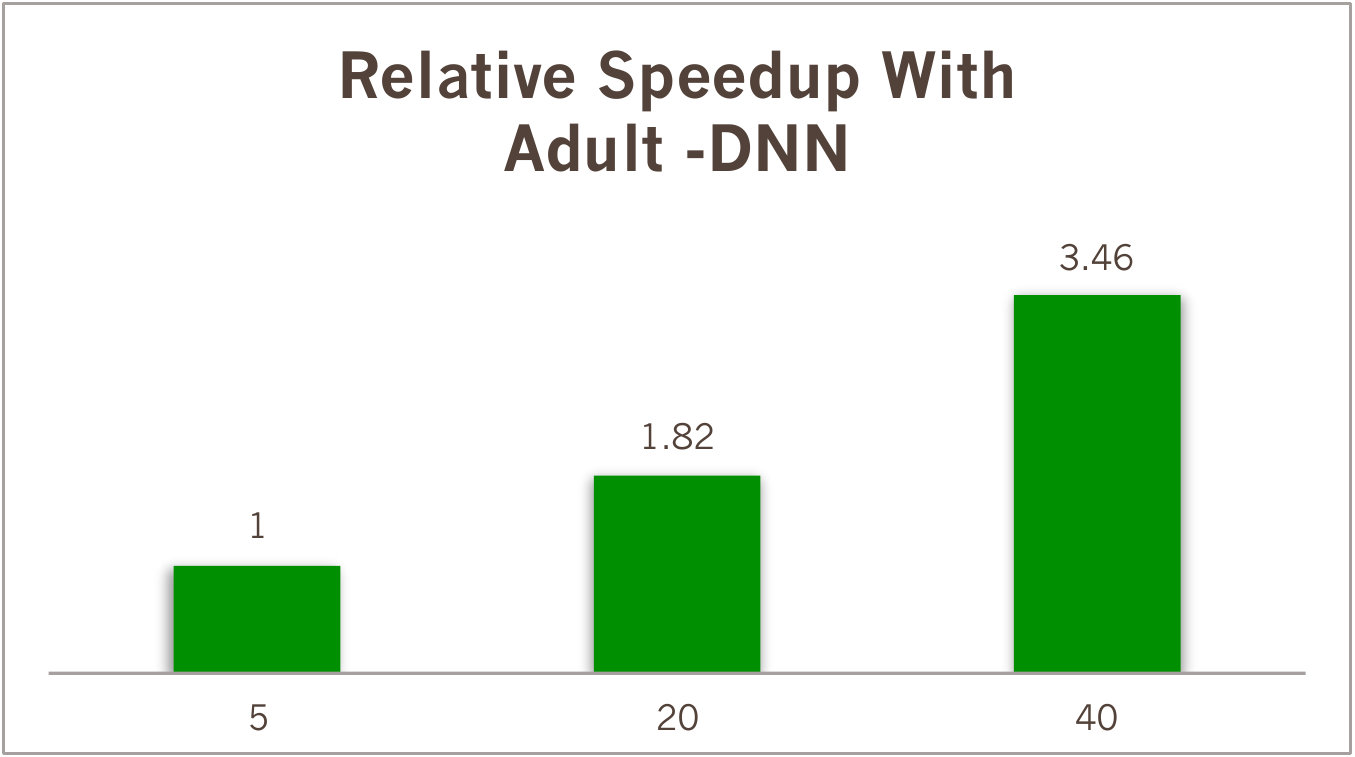}}
\caption{Relative Speedup to 5-core on Adult DNN using up to
40 cores}
\label{adult}
\end{center}
\vskip -0.2in
\end{figure}
The Adult data set classifies wither an adult makes more or less than $\$50,000$ a
year, based on variables like age, education, sex, native country, and
ten other variables.  The paper \cite{adultref} discusses the accuracy.
Figure~\ref{adult} shows the relative speedup in comparison to a 5-core
 evaluation. Similar to MNIST-DNN, we observe the benefits on each configuration. 

\subsection{Acoustic} \label{subsec:acoustic}
\begin{figure}[ht]
\vskip 0.2in
\begin{center}
\centerline{\includegraphics[width=\columnwidth]{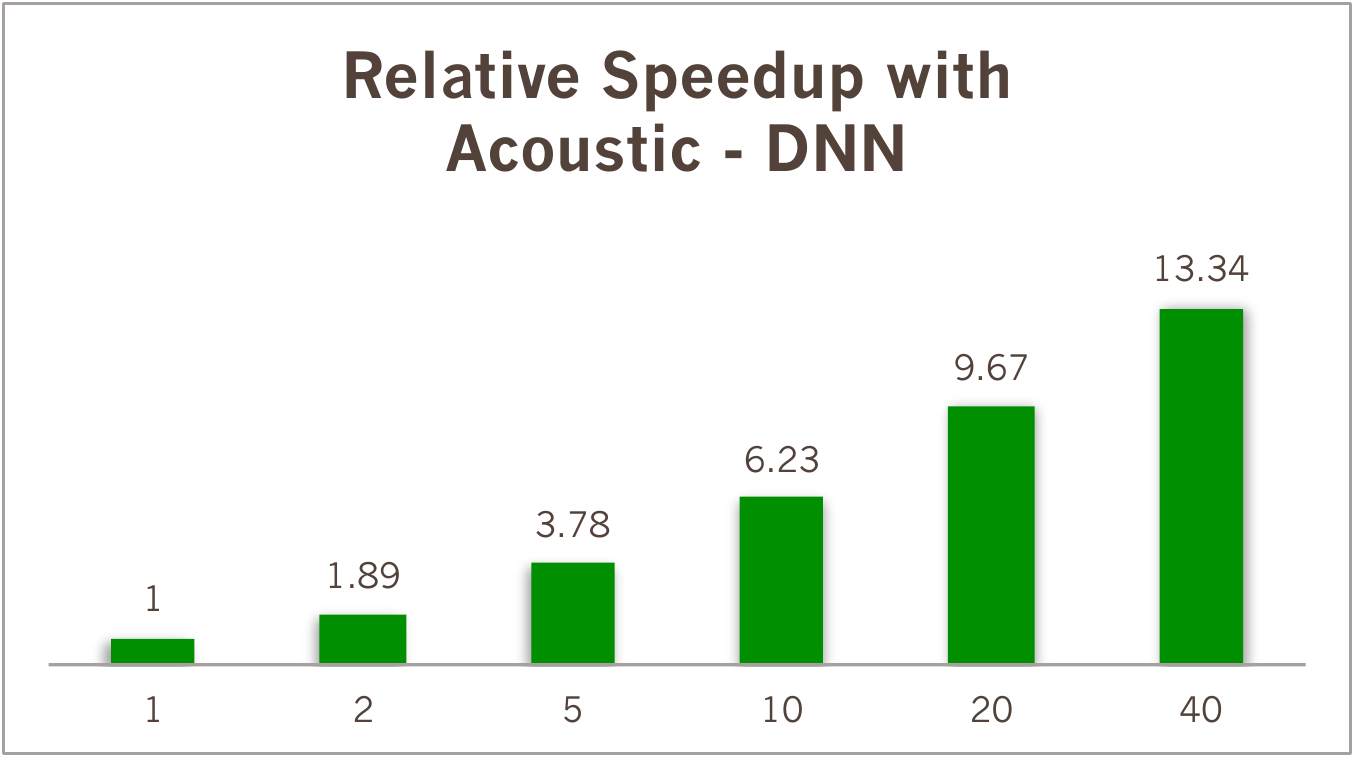}}
\caption{Relative Speedup to 1-core on Acoustic DNN using up to
40 cores}
\label{acoustic}
\end{center}
\vskip -0.2in
\end{figure}

Acoustic data set is used for vehicular classification in distributed
sensor networks. It has 78,823 samples, 3 classes and 50 features. 
Figure~\ref{acoustic} shows the relative speedup of the approach.

Similar to MNIST-DNN, we observe significant
speedup in comparison to the default implementation. We also observe
excellent scaling -- while we observe tapering at 32 cores, due to
reduced work per core. 

\subsection{CIFAR10} \label{subsec:CIFAR}
The CIFAR10 data set consist of $32\times 32$ pixel images with three
color channels.  The data set contains 50000 training images and 10000
testing images to be classified.  CIFAR10 consists of ten classes and
the examples are divided evenly among them. We consider a DNN and CNN
evaluation of the CIFAR10 data set.

\subsubsection{CIFAR10-DNN} 
\begin{figure}[ht]
\vskip 0.2in
\begin{center}
\centerline{\includegraphics[width=\columnwidth]{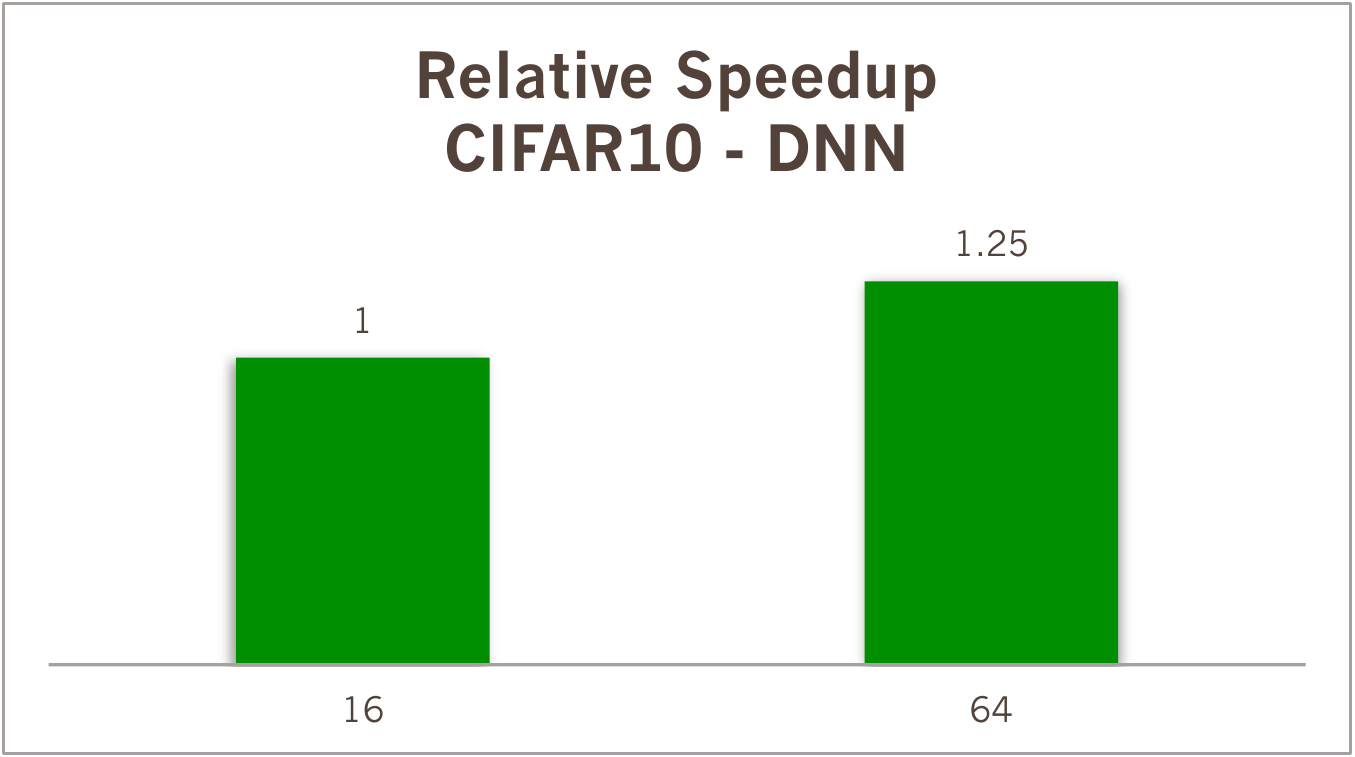}}
\caption{Relative Speedup to 16-core on CIFAR10-DNN using up to
64 cores}
\label{cifar10dnn}
\end{center}
\vskip -0.2in
\end{figure}

Figure~\ref{cifar10dnn} shows relative speedup for CIFAR10
using DNN. We observe a speedup of {\bf 2.97x} for 16 cores, and {\bf
3.37x} improvement with 64 cores. As expected, the parallel efficiency
decreases with strong scaling, as the work per core decreases. Since
CIFAR10 consists of images, it is usually evaluated with CNN. However,
even with a relatively modest size of the network, we were able to
easily achieve $\approx$ 55\% accuracy under an hour of training.

\subsubsection{CIFAR10-CNN} 
\begin{figure}[ht]
\vskip 0.2in
\begin{center}
\centerline{\includegraphics[width=\columnwidth]{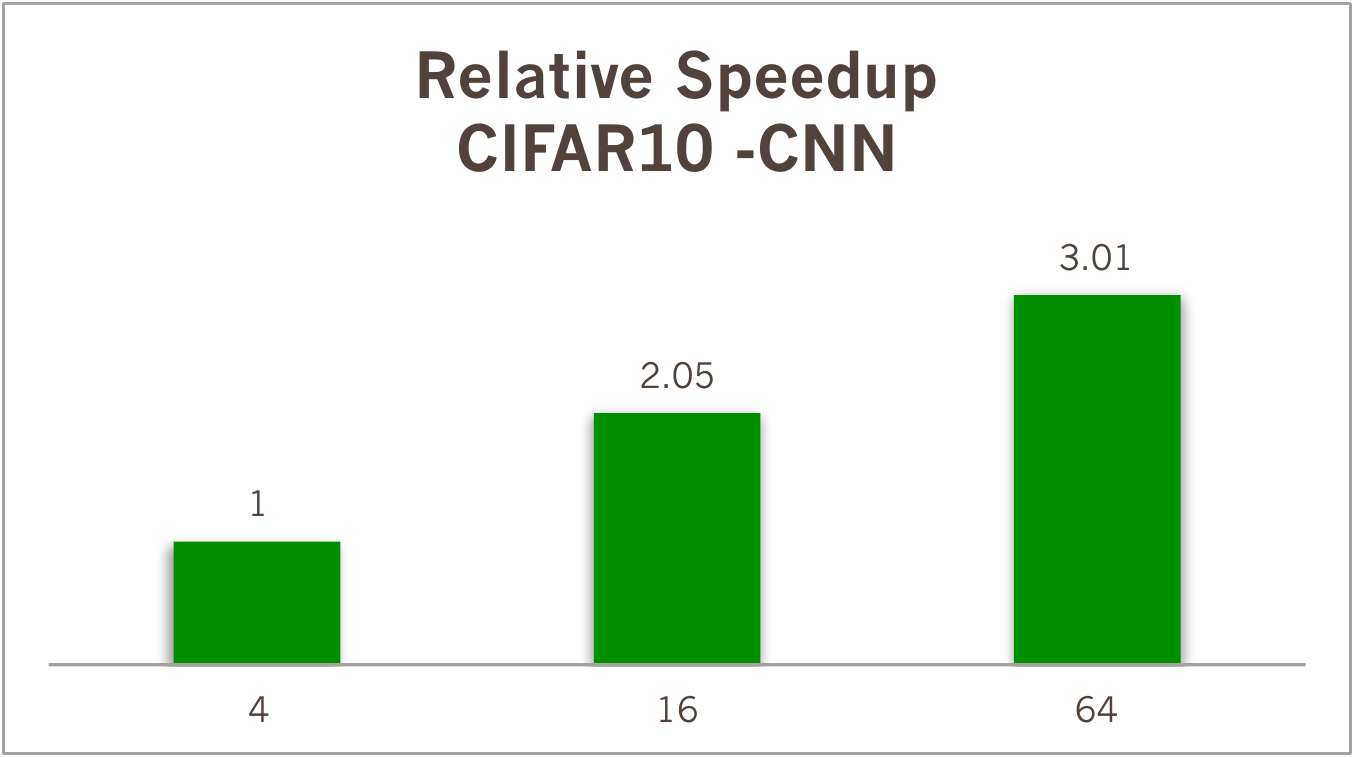}}
\caption{Relative Speedup to 4-core on CIFAR10-CNN using up to
64 cores}
\label{cifar10cnn}
\end{center}
\vskip -0.2in
\end{figure}

Figure~\ref{cifar10cnn} shows the relative speedup for
CIFAR10 using CNN. We observe that unlike the DNN case, the relative
improvements are modest.
\subsection{Higgs} \label{subsec:Higgs}
The last data set that we look at is the HIGGS data set.  It consists of
$11,000,000$ samples (the last $100,000$ of which are for testing)
generated by Monte Carlo simulations to emulate LHC data.  It has 28
features and every element is either classified as ``signal'' or
``background.''  It has been studied in \cite{NIPS2014_5351,
Baldi:2014kfa} with the goal of showing that deep learning has a place
in analyzing collider data in the search for new particles.

We evaluated HIGGS data set using up to 80 cores and achieved 2.6x
speedup in comparison to running on 20 cores.

\section{Related Work}
\label{sec:related}
Several researchers have conducted in-depth exploration of MLDM
algorithms, with a few focusing on scalability to multi-core/many-core
systems. A few researchers have considered execution on large scale
systems.

Several programming models have been proposed for large scale MLDM
algorithms. Mapreduce programming model provides large scale parallel
execution using Map and reduce tasks.  While original Mapreduce
programming model is generic, its actual incarnations (such as Hadoop)
have been widely critiqued for performance reasons. Recently proposed
programming models such as Spark, and associated MLDM libraries such as
MLlib support in-memory iterative MLDM algorithms. Other recent systems
include GraphLab -- which is primarily geared towards vertex based
computations for linked data structures. Similarly, MillWheel is used
for stream graph processing, but not necessarily suitable for large
scale MLDM algorithms.

Recently, several toolkits have become popular for MLDM algorithms.
Several MLDM toolkits which support sequential execution such as Weka,
Matlab, Scikit, Orange and libsvm have been very widely used for data
analysis. With recent developments in Deep Learning algorithms, several
implementations of Deep Learning algorithms have become available for
multi-core and many-core systems such as Theano, CuDNN and Caffe.

A few other toolkits support execution on large scale systems.  These
toolkits include Microsoft DMTK and Machine Learning Toolkit for Extreme
Scale (MaTEx).  Recently released TensorFlow supports MLDM algorithms
with automatic differentiation. It is readily available for deployment
with multi-core and many-core clusters. It contains several
optimizations such as Adaptive Gradient Descent (AdaGrad), Dropout for
regularization among others. However, it is not ready for large scale
systems -- such as executions on clusters.  

\section{Conclusions}
\label{sec:conclusions}
In this paper, we have proposed a design to alleviate the distributed memory limitations of
TensorFlow. 
We have considered several programming models,
especially Mapreduce based programming models (Hadoop, and Spark) and
conclude that neither of them are geared towards realizing the peak
potential of system, while TensorFlow is geared towards exploiting the
architecture effectively using C++ backend and state of the art linear
algebra packages. We have used Message Passing Interface (MPI)
as the communication interface for parallelizing TensorFlow on
distributed memory subsystems. We have specified the changes which were
requires to realize the implementation on distributed memory systems.
Specifically, we conclude that these changes are minimal and require
no changes to the TensorFlow runtime! Our evaluation of the proposed
extensions with several well known datasets such as MNIST, CIFAR-10,
Adult and Higgs reveals the performance efficiency of the proposed
implementation.

\section{Acknowledgement}
This research is conducted under Analysis in Motion (AIM) Laboratory
Directed Research and Development (LDRD) Initiative.

\bibliographystyle{IEEEtran}
\bibliography{IEEEabrv,vishnu,vishnu_ml,FasterLearning,extra,psmo}
\end{document}